\documentclass[pra,twocolumn,amsmath,amssymb,superscriptaddress]{revtex4-1}
\usepackage{epsfig,amsmath}
\usepackage{subfigure}
\usepackage{graphicx}
\usepackage{dcolumn}
\usepackage{stmaryrd}
\usepackage{mathrsfs}
\usepackage{pifont}
\usepackage{amsthm}
\usepackage{amssymb}
\usepackage{bm}
\usepackage{latexsym}
\usepackage[colorlinks=true,linkcolor=blue,citecolor=blue]{hyperref}
\usepackage{color}
\usepackage{epstopdf}

\usepackage{booktabs}
\usepackage{threeparttable}
\usepackage{multirow}

\begin{document}

\title{Universal quantum control by non-Hermitian Hamiltonian}

\author{Zhu-yao Jin}
\affiliation{School of Physics, Zhejiang University, Hangzhou 310027, Zhejiang, China}

\author{Jun Jing}
\email{Contact author: jingjun@zju.edu.cn}
\affiliation{School of Physics, Zhejiang University, Hangzhou 310027, Zhejiang, China}

\date{\today}

\begin{abstract}
Conventional manipulations over quantum systems for such as coherent population trapping and unidirectional transfer focus on Hamiltonian engineering while regarding the system's manifold geometry and constraint equation as secondary causes. Here we treat them on equal footing in controlling a finite-dimensional quantum system under a time-dependent non-Hermitian Hamiltonian, which is inspired by the D'Alembert principle of regarding active force, constraint force, and inertial force in an unbiased way. Under the biorthogonal condition, the non-Hermitian Hamiltonian could be triangularized in a constraint picture spanned by a set of completed and orthonormal basis states, which is found to be a sufficient condition to construct at least one universal nonadiabatic passage in both bra and ket spaces. The passage ends up with a desired target state that is automatically normalized without artificial normalization in the existing treatments for non-Hermitian quantum systems. Moreover, the passage is found to be robust against the parametric deviation when the real part of its global phase is rapidly varying with time. Our protocol is explicitly verified for the perfect population transfer in the two-level system and the chiral population transfer in the three-level system. It generalizes our framework of universal quantum control to the field of the biorthogonal quantum mechanics.
\end{abstract}

\maketitle

\section{Introduction}

Dynamics of quantum system under control is generally described by the time-dependent Schr\"odinger equation~\cite{Schrodinger1926AnUndulatory}. The prior investigations over both Hermitian and non-Hermitian quantum mechanics~\cite{Siegert1939Onthe,Herman1958Unified,Herman1962Unified} focused primarily on the properties of system Hamiltonian, e.g., the energy spectra and the instantaneous eigenstates, while regarding the manifold geometry and the boundary constraints as minor elements. Distinctively, the present work attempts to treat them on equal footing in the control of finite-dimensional systems under a time-dependent non-Hermitian Hamiltonian~\cite{Siegert1939Onthe,Herman1958Unified,Herman1962Unified}.

The fundamentals of quantum mechanics are built upon the Hermitian Hamiltonian and the Hermiticity guarantees the probability conservation and real-valued energy spectra. However, open quantum systems typically exhibit undesirable flows of energy, particles, and information between the designated Hilbert space and the external space~\cite{Ashida2020NonHermitian}, yielding non-Hermitian revision to the system Hamiltonian~\cite{Ashida2020NonHermitian,Bender2024PTsymmetric}. The non-Hermitian Hamiltonian is ubiquitous and has recently attracted growing attention in many platforms~\cite{Zeuner2015Observation,ElGanainy2018Nonhermitian,Wang2022Giant,Li2019Observation,Takasu2020PTsymmetric,
Liang2022Dynamic,Bender2013Observation,Yoshida2019Exceptional,Stehmann2004Observation,Schindler2011Experimental,
Zhang2017Observation,Harder2018Level,Wang2019Nonreciprocity,Yang2020Unconventional,Qian2024probing,Wang2024Enhancement,
Hong2024Synchronization}, such as optical waveguide systems~\cite{Zeuner2015Observation,ElGanainy2018Nonhermitian,Wang2022Giant}, cold atoms~\cite{Li2019Observation,Takasu2020PTsymmetric,Liang2022Dynamic}, mechanical systems~\cite{Bender2013Observation,Yoshida2019Exceptional}, electronic circuits~\cite{Stehmann2004Observation,Schindler2011Experimental}, and cavity-electromagnonical systems~\cite{Zhang2017Observation,Harder2018Level,Wang2019Nonreciprocity,Yang2020Unconventional,
Qian2024probing,Wang2024Enhancement}.

As the counterpart to the critical points~\cite{Sachdev1999Quantumphase} in the Hermitian systems~\cite{Kawabata2017Information}, the exceptional points (EPs)~\cite{Heiss2000Repulsion,Dembowski2001Experimental,Heiss2012Exceptional,Hanai2020Critical,Zhang2017Observation}, where both eigenvalues and their corresponding eigenvectors coalesce, of the non-Hermitian systems can be approached by tuning the system parameters. In most studies, the spectral singularity of the non-Hermitian Hamiltonian emerges around EPs~\cite{Heiss2012Exceptional}, resulting in various unconventional phenomena, such as unidirectional invisibility~\cite{Lin2011Unidirectional,Feng2013Experimental,Yin2013Unidirectional}, coherent perfect absorber~\cite{Longhi2010PTsymmetric,Chong2010Coherent,Sun2014Experimental}, enhanced sensitivity~\cite{Hodaei2017Enhanced,Chen2016PTsymmetry,Liu2016Metrology}, nonreciprocal light propagation~\cite{Peng2014PTsymmetric,Ramerani2010Unidirectional,Chang2014Paritytime}, and suppression and revival of single-mode lasing~\cite{Peng2014Lossinduced,Brandstetter2014Reversing}.

On the other hand, many state-transfer protocols, including the rapid adiabatic passage~\cite{Feilhauer2020Encircling}, the counterdiabatic driving methods~\cite{Ibanez2011Shortcuts,Torosov2013NonHermitian,Torosov2014NonHermitian,Chen2016Method,Luan2022Shortcuts,
Zhang2022NonHermitian}, and the Lewis-Riesenfeld theory for invariants~\cite{Ibanez2011Shortcuts,Luo2015Dynamical}, have been explored in the non-Hermitian two-level system~\cite{Bender2007Faster,Ibanez2011Shortcuts,Torosov2013NonHermitian,Chen2016Method,Guery2019Shortcuts,
Luan2022Shortcuts,Feilhauer2020Encircling}, three-level system~\cite{Torosov2014NonHermitian,Guery2019Shortcuts}, and cavity-electromagnonical system~\cite{Zhang2022NonHermitian}. Currently, these non-Hermitian protocols are limited to the parity-time (PT) symmetrical Hamiltonian in an effective two-dimensional subspace~\cite{Ibanez2011Shortcuts,Torosov2013NonHermitian,Torosov2014NonHermitian,Chen2016Method,Luan2022Shortcuts,
Zhang2022NonHermitian} without crossing EPs. They rely on complex laser interactions or time-dependent gain and loss rates of levels and modes~\cite{Ibanez2011Shortcuts,Torosov2013NonHermitian,Torosov2014NonHermitian,Chen2016Method,Luan2022Shortcuts,
Zhang2022NonHermitian}, which are challenging in experimental implementation. It was shown that under a non-Hermitian driving, an open two-level system can be engineered from the ground state to a specific target state with an unlimited speed~\cite{Bender2007Faster}, which is in sharp contrast to that under a Hermitian Hamiltonian. The control over non-Hermitian systems exhibits counterintuitive behaviors due to the biorthogonality and probability nonconservation~\cite{Ibanez2014Adiabaticity,Ashida2020NonHermitian}. For example, the adiabatic bidirectional state transfer in optical waveguides will become a unidirectional transfer when a loss occurs in one of the waveguides~\cite{Graefe2013Breakdown,Doppler2016Dynamically}. Also, the unit-fidelity transfer in non-Hermitian systems remains severely constrained unless by artificial normalization~\cite{Feilhauer2020Encircling,Ibanez2011Shortcuts,Torosov2014NonHermitian,Luan2022Shortcuts}. A unified theoretical framework that treats all possible time-dependent Hamiltonians, either Hermitian or non-Hermitian (avoiding EPs), on equal footing is desired for the control over general quantum systems.

In this paper, we derive a general theory to solve the time-dependent Schr\"odinger equation for the finite-dimensional systems controlled by the non-Hermitian biorthogonal Hamiltonian~\cite{Brody2013Biorhogonal}, as a substantial extension of universal quantum control framework proposed for Hermitian quantum mechanics~\cite{Jin2025Universal,Jin2025Entangling,Jin2025ErrCorr,Jin2025Rydberg}. At the operational level, the time-dependent and non-Hermitian Hamiltonian can be triangularized in a rotated picture. It renders at least one transitionless passage in both bra and ket spaces, along which the system can deterministically evolve to a desired target state. The system state can be automatically normalized at the end of the controlled passage without any artificial normalization, in sharp contrast to the conventional treatments for non-Hermitian systems~\cite{Ibanez2014Adiabaticity,Daley2009Atomic,Uzdin2012Timedependent,Ashida2018Full-Counting,
Dora2020Quantum,Motta2020Determining}. Moreover, the passage is found to be insensitive to the parametric deviation when the real part of global phase is manipulated with a fast rate. In addition to the perfect population transfer in a two-level system, the versatility of our theory is demonstrated with the chiral population transfers of both clockwise and counterclockwise styles in a three-level system in the presence of gain and loss.

The rest of this paper is structured as follows. In Sec.~\ref{general}, we introduce a general theoretical framework for solving the time-dependent Schr\"odinger equation of the finite-dimensional systems under an arbitrary biorthogonal Hamiltonian. The biorthogonal Hamiltonian can be triangularized in the ancillary picture and it suffices to activate at least one ancillary basis state as the useful universal passage in both bra and ket spaces. Section~\ref{NonHermTwo} exemplifies the passage-construction protocol for a paradigmatic two-level system. The robustness of passage can be enhanced by our correction mechanism. Section~\ref{NonHermThree} extends the protocol in a hybrid way to a three-level system and shows the chiral population transfer. In Sec.~\ref{conclusion}, we summarize the whole work. Appendix~\ref{dAlembert} briefly recalls the universal quantum control framework, in which the passage is analog to the virtual displacement in the D'Alembert principle. Appendix~\ref{appendix} shows the reduction from the triangularization of a non-Hermitian Hamiltonian to the diagonalization of a Hermitian Hamiltonian.

\section{Theoretical framework}\label{general}

Our study is conducted on an arbitrary $K$-dimensional system controlled by a time-dependent non-Hermitian Hamiltonian $H(t)$, i.e., $H(t)\neq H^\dagger(t)$. Under the assumption that the system dynamics avoids crossing EPs, the eigenstates in the bra and ket spaces are biorthogonal and completed~\cite{Brody2013Biorhogonal}. The system dynamics can be described by the following two groups of time-dependent Schr\"odinger equations as ($\hbar\equiv1$)
\begin{subequations}\label{SchEff}
\begin{align}
&i\frac{d|\psi_m(t)\rangle}{dt}=H(t)|\psi_m(t)\rangle,\label{SchHam}\\
&i\frac{d|\phi_m(t)\rangle}{dt}=H^\dagger(t)|\phi_m(t)\rangle,\label{SchHamNon}
\end{align}
\end{subequations}
with $1\leq m\leq K$, where $|\psi_m(t)\rangle$'s and $\langle\phi_m(t)|$'s are the pure-state solutions in the ket and bra spaces, respectively. They satisfy the biorthogonal relation as $\langle\phi_k(t)|\psi_m(t)\rangle=\delta_{km}$~\cite{Muga2004Complex,Ibanez2011Shortcuts} and the completed condition as $\sum_{m=1}^K|\psi_m(t)\rangle\langle\phi_m(t)|=\sum_{m=1}^K|\phi_m(t)\rangle\langle\psi_m(t)|=\mathcal{I}$. According to Eqs.~(\ref{SchHam}) and (\ref{SchHamNon}), we have
\begin{subequations}\label{Uform}
\begin{align}
|\psi_m(t)\rangle&=U_0(t)|\psi_m(t_0)\rangle,\label{U0form}\\
|\phi_m(t)\rangle&=V_0(t)|\phi_m(t_0)\rangle\label{U0formhat},
\end{align}
\end{subequations}
where $U_0(t)$ and $V_0(t)$ are the time-evolution operators in the ket and bra spaces, respectively. The initial state of the system can be represented by either $|\psi_m(t_0)\rangle$ or $\langle\phi_m(t_0)|$. Solving the time-dependent Schr\"odinger equation is generally challenging even for the standard Hermitian systems~\cite{Jin2025Universal,Jin2025Entangling,Jin2025ErrCorr,Jin2025Rydberg}, unless the pure-state solutions have already constituted a completed and orthonormal set for the Hilbert space of the system. It will become harder for a non-Hermitian Hamiltonian.

In the regime of Hermitian mechanics, the challenge of dynamics can be partially or completely addressed by our theory of universal quantum control~\cite{Jin2025Universal,Jin2025Entangling,Jin2025ErrCorr,Jin2025Rydberg}, which is inspired by the D'Alembert principle about the virtual displacement determined by active, constraint, and inertial forces (see Appendix~\ref{dAlembert} for details). This theory can be reduced to various time-modulated controls under respective conditions, including the stimulated Raman adiabatic passage, the nonadiabatic holonomic transformation, the Lewis-Riesenfeld theory for invariants, and counterdiabatic driving methods. In the following, it will be extended to the regime of non-Hermitian Hamiltonian avoiding EPs, irrespective of PT symmetry.

Similar to the universal quantum control~\cite{Jin2025Universal,Jin2025Entangling,Jin2025ErrCorr,Jin2025Rydberg} performed in a rotated picture, we are working in an ancillary picture for the non-Hermitian system, which is spanned by the basis states $|\mu_k(t)\rangle$'s, $1\leq k\leq K$, with $\langle\mu_k(t)|\mu_m(t)\rangle=\delta_{km}$. They span the Hilbert space of system by $\sum_{k=1}^K|\mu_k(t)\rangle\langle\mu_k(t)|=\mathcal{I}$. A brief recipe of constructing $|\mu_k(t)\rangle$ was provided for general two-band systems~\cite{Jin2025Entangling}. In the rotating frame with respect to $\mathcal{V}(t)\equiv\sum_{k=1}^{K}|\mu_k(t)\rangle\langle\mu_k(t_0)|$, the system Hamiltonian $H(t)$ can be rotated as
\begin{equation}\label{Hamrot}
\begin{aligned}
&H_{\rm rot}(t)=\mathcal{V}^\dagger(t)H(t)\mathcal{V}(t)-i\mathcal{V}^\dagger(t)\frac{d}{dt}\mathcal{V}(t)\\
=&\sum_{k=1}^{K}\sum_{m=1}^K\left[\mathcal{H}_{km}(t)-\mathcal{A}_{km}(t)\right]|\mu_k(t_0)\rangle\langle\mu_m(t_0)|\\
\equiv&\mathcal{V}^\dagger(t)\left[\mathcal{H}(t)-\mathcal{A}(t)\right]\mathcal{V}(t),
\end{aligned}
\end{equation}
where the dynamical term $\mathcal{H}_{km}(t)\equiv\langle\mu_k(t)|H(t)|\mu_m(t)\rangle$ and the gauge potential $\mathcal{A}_{km}\equiv i\langle\mu_k(t)|\dot{\mu}_m(t)\rangle$~\cite{Michael2017Geometry,Claeys2019Floquet,Takahashi2024Shortcuts} are the elements in the $k$th row and $m$th column of the non-Hermitian matrix $\mathcal{H}(t)$ and the Hermitian matrix $\mathcal{A}(t)$, respectively. As for the Hamiltonian $H^\dagger(t)$ in Eq.~(\ref{SchHamNon}), it can be rotated as the Hermitian conjugate of $H_{\rm rot}(t)$, i.e., $H_{\rm rot}^\dagger(t)\equiv\mathcal{V}^\dagger(t)[\mathcal{H}^\dagger(t)-\mathcal{A}(t)]\mathcal{V}(t)$. Note the matrix $\mathcal{A}(t)$ embodies a purely geometric structure, being uniquely determined by the intrinsic properties of the differential manifolds of the ancillary picture. In fact by Eq.~(\ref{Hamrot}), $\mathcal{A}(t)$ is determined by the constraint equation $\mathcal{A}(t)|\mu_k(t)\rangle=id|\mu_k(t)\rangle/dt$ with the ancillary basis state $|\mu_k(t)\rangle$.

Consequently, the time-dependent Schr\"odinger equations~(\ref{SchHam}) and (\ref{SchHamNon}) are transformed to be
\begin{subequations}\label{Schrot}
\begin{align}
i\frac{d|\psi_m(t)\rangle_{\rm rot}}{dt}&=H_{\rm rot}(t)|\psi_m(t)\rangle_{\rm rot},\label{SchHamRot}\\
i\frac{d|\phi_m(t)\rangle_{\rm rot}}{dt}&=H_{\rm rot}^\dagger(t)|\phi_m(t)\rangle_{\rm rot},\label{SchNonHamRot}
\end{align}
\end{subequations}
with the rotated pure-state solutions
\begin{equation}\label{relate}
\begin{aligned}
& |\psi_m(t)\rangle_{\rm rot}=\mathcal{V}^\dagger(t)|\psi_m(t)\rangle, \\
& |\phi_m(t)\rangle_{\rm rot}=\mathcal{V}^\dagger(t)|\phi_m(t)\rangle.
\end{aligned}
\end{equation}
The relevant time-evolution operators for Eqs.~(\ref{SchHamRot}) and (\ref{SchNonHamRot}) can be written in the Dyson series~\cite{Dyson1949TheRadiation} as
\begin{equation}\label{UrotDyson}
\begin{aligned}
&U_{\rm rot}(t)=\hat{T}e^{-i\int_{t_0}^tH_{\rm rot}(t')dt'}\\
=&\sum_{n=0}^{\infty}(-i)^n\int_{t_0}^tdt_1\cdots\int_{t_0}^{t_{n-1}}dt_nH_{\rm rot}(t_1)\cdots H_{\rm rot}(t_n),
\end{aligned}
\end{equation}
and
\begin{equation}\label{VrotDyson}
\begin{aligned}
&V_{\rm rot}(t)=\hat{T}e^{-i\int_{t_0}^tH_{\rm rot}^\dagger(t')dt'}\\
=&\sum_{n=0}^{\infty}(-i)^n\int_{t_0}^tdt_1\cdots\int_{t_0}^{t_{n-1}}dt_nH_{\rm rot}^\dagger(t_1)\cdots H_{\rm rot}^\dagger(t_n),
\end{aligned}
\end{equation}
respectively, where $\hat{T}$ is the time-ordering operator and the zeroth-order term is the identical operator.

It is generally hard to obtain the solution of $U_{\rm rot}(t)$ or $V_{\rm rot}(t)$ in a closed form due to the noncommutativity of the time-dependent Hamiltonian $H_{\rm rot}(t)$ at distinct time points. In case of the Hermitian Hamiltonian~\cite{Jin2025Universal,Jin2025Entangling,Jin2025ErrCorr,Jin2025Rydberg}, $H_{\rm rot}(t)=H_{\rm rot}^\dagger(t)$ can be partially or completely diagonalized by the von Neumann equation about the projection operator $\Pi_k(t)\equiv|\mu_k(t)\rangle\langle\mu_k(t)|$. Then the obtained parametric conditions for $H(t)$ can activate the relevant ancillary basis state $|\mu_k(t)\rangle$ to be the universal passage and the time-evolution operator can be partially or completely determined (see Appendix~\ref{dAlembert} for details). However, the non-Hermitian characteristic of the Hamiltonian, i.e., $\mathcal{H}_{km}(t)\ne\mathcal{H}_{mk}^*(t)$, $k\ne m$, can cause $H_{\rm rot}(t)$ in Eq.~(\ref{Hamrot}) and its Hermitian conjugate to be nondiagonalizable in any picture. It means that the von Neumann equation should be generalized to adapt to the non-Hermitian Hamiltonian.

\subsection{Main result}

We prove that the lower triangularization of the rotated Hamiltonian $H_{\rm rot}(t)$ in Eq.~(\ref{Hamrot}), i.e., $\mathcal{H}_{km}(t)-\mathcal{A}_{km}(t)=0$ with $k<m$, is a sufficient condition for activating one ancillary basis state as the universal passage, determined by $U_{\rm rot}(t)$. In the ancillary picture, the lower triangularization is equivalent to
\begin{equation}\label{Tri}
\Pi_k(t)\left[\mathcal{H}(t)-\mathcal{A}(t)\right]\sum_{k'>k}^K\Pi_{k'}(t)=0, \quad 1\leq k<K.
\end{equation}
Equations~(\ref{Hamrot}) and (\ref{Tri}) are mathematically grounded in Schur's theorem~\cite{Sheldon2023Linear}, which guarantees that any finite-dimensional complex matrix $\mathcal{H}(t)-\mathcal{A}(t)$ can be triangularized via a unitary transformation, such as $\mathcal{V}(t)$. Specially when $\mathcal{H}(t)-\mathcal{A}(t)$ is Hermitian~\cite{Jin2025Universal}, Schur's theorem is simplified as the spectral theorem~\cite{Sheldon2023Linear} for the unitary diagonalizability and accordingly, Eq.~(\ref{Tri}) reduces to the von Neumann equation~(\ref{von}). The details can be found in Appendix~\ref{appendix}.

\subsection{Sufficient condition towards solvable time-evolution operator}

If the rotated Hamiltonian $H_{\rm rot}(t)$ in Eq.~(\ref{Hamrot}) has been lower triangularized under Eq.~(\ref{Tri}), then it can be written as
\begin{equation}\label{HamLow}
H_{\rm rot}(t)=\sum_{k\ge m}^K\sum_{m=1}^K\left[\mathcal{H}_{km}(t)-\mathcal{A}_{km}(t)\right]|\mu_k(t_0)\rangle\langle\mu_m(t_0)|.
\end{equation}
Its integral over time is
\begin{equation}\label{Hamint}
\begin{aligned}
& \int_{t_0}^tH_{\rm rot}(t')dt'=\sum_{k\ge m}^K\sum_{m=1}^Kf_{km}(t)|\mu_k(t_0)\rangle\langle\mu_m(t_0)|,\\
& f_{km}(t)\equiv\int_{t_0}^t\left[\mathcal{H}_{km}(t')-\mathcal{A}_{km}(t')\right]dt'.
\end{aligned}
\end{equation}
Subsequently, the $n$th-order term in Eq.~(\ref{UrotDyson}) is
\begin{equation}\label{HamintNpower}
\begin{aligned}
&(-i)^n\int_{t_0}^tdt_1\cdots\int_{t_0}^{t_{n-1}}dt_nH_{\rm rot}(t_1)\cdots H_{\rm rot}(t_n)\\
&=\sum_{m=1}^K\frac{(-i)^n}{n!}[f_{mm}(t)]^n|\mu_m(t_0)\rangle\langle\mu_m(t_0)|\\
&+\sum_{k>m}^K\sum_{m=1}^{K-1}\tilde{f}_{km}(t)|\mu_k(t_0)\rangle\langle\mu_m(t_0)|.
\end{aligned}
\end{equation}
Equations~(\ref{UrotDyson}) and (\ref{HamintNpower}) yield a formal solution to the evolution operator:
\begin{equation}\label{Urot}
\begin{aligned}
U_{\rm rot}(t)&=\sum_{m=1}^Ke^{-if_{mm}(t)}|\mu_m(t_0)\rangle\langle\mu_m(t_0)|\\
&+\sum_{k>m}^K\sum_{m=1}^{K-1}u_{km}(t)|\mu_k(t_0)\rangle\langle\mu_m(t_0)|,
\end{aligned}
\end{equation}
where the off-diagonal elements $u_{km}(t)$ as well as $\tilde{f}_{km}(t)$ in Eq.~(\ref{HamintNpower}) are hardly obtained in general and yet not relevant to our theory. If the system is initially populated on the state $|\mu_K(t_0)\rangle$, then the time-evolution operator in Eq.~(\ref{Urot}) is equivalent to
\begin{equation}\label{UrotOne}
U_{\rm rot}(t)=e^{-if_{KK}(t)}|\mu_K(t_0)\rangle\langle\mu_K(t_0)|.
\end{equation}
Rotating back to the original picture implied by Eq.~(\ref{relate}), we have
\begin{equation}\label{U0}
U_0(t)=\mathcal{V}(t)U_{\rm rot}(t)=e^{-if_{KK}(t)}|\mu_K(t)\rangle\langle\mu_K(t_0)|.
\end{equation}
It suggests that the conditions of lower triangularizing $H_{\rm rot}(t)$ in Eq.~(\ref{HamLow}) can activate the ancillary base $|\mu_{k=K}(t)\rangle$ to be a useful passage in the ket space of the non-Hermitian system. When the system starts the passage $|\mu_K(t_0)\rangle$ from an arbitrary point $t_0$, it evolves to the target state $|\mu_K(t)\rangle$ with an accumulated global phase $f_{KK}(t)$.

The lower triangularization of $H_{\rm rot}(t)$ in Eq.~(\ref{HamLow}) renders the upper triangularization of $H_{\rm rot}^\dagger(t)$ as
\begin{equation}\label{HamUpper}
H_{\rm rot}^\dagger(t)=\sum_{k\ge m}^K\sum_{m=1}^K\left[\mathcal{H}^*_{km}(t)-\mathcal{A}_{km}(t)\right]|\mu_m(t_0)\rangle\langle\mu_k(t_0)|,
\end{equation}
and its time integral is
\begin{equation}\label{HamUpperInt}
\int_{t_0}^tH_{\rm rot}^\dagger(t')dt'=\sum_{k\ge m}^K\sum_{m=1}^Kf_{km}^*(t)|\mu_m(t_0)\rangle\langle\mu_k(t_0)|.
\end{equation}
Similarly, the time-evolution operator $V_{\rm rot}(t)$ in the bra space can be written as
\begin{equation}\label{Urotconjud}
\begin{aligned}
V_{\rm rot}(t)&=\sum_{m=1}^Ke^{-if_{mm}^*(t)}|\mu_m(t_0)\rangle\langle\mu_m(t_0)|\\
&+\sum_{k>m}^K\sum_{m=1}^{K-1}u_{km}^*(t)|\mu_m(t_0)\rangle\langle\mu_k(t_0)|.
\end{aligned}
\end{equation}
If the system is found to be at $|\mu_1(t_0)\rangle$ for arbitrary $t_0$, $V_{\rm rot}(t)$ can be reduced to
\begin{equation}\label{Urotadj}
V_{\rm rot}(t)=e^{-if_{11}^*(t)}|\mu_1(t_0)\rangle\langle\mu_1(t_0)|.
\end{equation}
Rotating back to the original picture, the time-evolution operator can be written as
\begin{equation}\label{U0adj}
V_0(t)=\mathcal{V}(t)V_{\rm rot}(t)=e^{-if_{11}^*(t)}|\mu_1(t)\rangle\langle\mu_1(t_0)|.
\end{equation}
It means that under the same condition in Eq.~(\ref{Tri}), another ancillary basis state $\langle\mu_{k=1}(t)|$ can be activated as a universal passage in the bra (dual) space of the same non-Hermitian system. Note $V_{\rm rot}(t)$ and $V_0(t)$ are the expressions in the ket space corresponding to the practical operators in the bra space.

\section{State transfer in non-Hermitian two-level systems}\label{NonHermTwo}

\begin{figure}[htbp]
\centering
\includegraphics[width=0.4\linewidth]{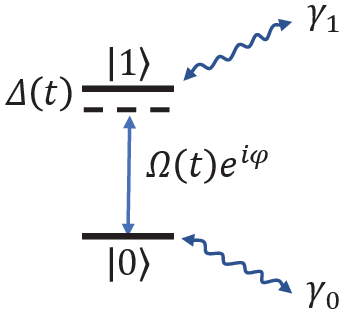}
\caption{Sketch of an open two-level quantum system under control. The transition $|0\rangle\leftrightarrow|1\rangle$ is driven by the off-resonant laser field with the detuning $\Delta(t)$, the Rabi frequency $\Omega(t)$, and the initial phase $\varphi$. The gain or loss rates $\gamma_0$ and $\gamma_1$ are associated with the levels $|0\rangle$ and $|1\rangle$, respectively, due to their coupling with the environment. }\label{model}
\end{figure}

In this section, the universal control framework for the non-Hermitian quantum system is examined in an open two-level system, that is composed of the ground state $|0\rangle$ and the excited state $|1\rangle$ as depicted in Fig.~\ref{model}. The transition $|0\rangle\leftrightarrow|1\rangle$ is driven by an off-resonant laser field with the detuning $\Delta(t)$, the Rabi frequency $\Omega(t)$, and the initial phase $\varphi$. In addition, the levels $|0\rangle$ and $|1\rangle$ are assumed with the gain or loss rate $\gamma_0$ and $\gamma_1$~\cite{Streed2006Continuous,Graefe2008Meanfield,Graefe2010Quantum,Ibanez2011Interaction,
Moiseyev2011Crossing,ElGanainy2012Local,Xiao2012Coherent}, respectively. $\gamma_0$ and $\gamma_1$ can be time independent or time dependent in theory. In what follows, they are assumed to be time independent unless otherwise stated. In general, the whole Hamiltonian reads
\begin{equation}\label{Ham}
\begin{aligned}
&H(t)=\frac{\Delta(t)}{2}(|1\rangle\langle1|-|0\rangle\langle0|)+\frac{1}{2}\Big[e^{i\xi_0}\gamma_0|0\rangle\langle0|\\
+&e^{i\xi_1}\gamma_1|1\rangle\langle1|\Big]+\left[\frac{1}{2}\Omega(t)e^{i\varphi}|1\rangle\langle 0|+{\rm H.c.}\right],
\end{aligned}
\end{equation}
where the gain or loss effect of the levels depends on the phases $\xi_0$ and $\xi_1$. The non-Hermitian Hamiltonian~(\ref{Ham}) with $|\xi_0|=|\xi_1|$ can be verified to stay at the unbroken and broken phases about PT symmetry~\cite{Ashida2020NonHermitian,Bender2024PTsymmetric} when $\Delta(t)=0$ and $\Delta(t)\ne0$, respectively. Our theory is applicable to both of them.

By virtue of the universal quantum control~\cite{Jin2025Universal,Jin2025Entangling,Jin2025ErrCorr,Jin2025Rydberg}, the two-level system dynamics under any Hamiltonian can be described in the following rotated picture, whose basis states can be chosen as
\begin{equation}\label{AnciTwo}
\begin{aligned}
|\mu_1(t)\rangle&=\cos\theta(t)e^{i\frac{\alpha(t)}{2}}|0\rangle-\sin\theta(t)e^{-i\frac{\alpha(t)}{2}}|1\rangle,\\
|\mu_2(t)\rangle&=\sin\theta(t)e^{i\frac{\alpha(t)}{2}}|0\rangle+\cos\theta(t)e^{-i\frac{\alpha(t)}{2}}|1\rangle,
\end{aligned}
\end{equation}
where the parameters $\theta(t)$ and $\alpha(t)$ manipulate the population and the local phase of the states, respectively. Substituting them to the triangularization equation~(\ref{Tri}) with the non-Hermitian Hamiltonian~(\ref{Ham}), we have
\begin{equation}\label{ConditionTwoGene}
\begin{aligned}
\Omega(t)&=\frac{-4\dot{\theta}(t)+(\gamma_0\sin\xi_0-\gamma_1\sin\xi_1)\sin2\theta(t)}{2\sin\left[\varphi+\alpha(t)\right]},\\
\Delta(t)&=\dot{\alpha}(t)+\Omega(t)\cot2\theta(t)\cos\left[\varphi+\alpha(t)\right]\\
&+\frac{1}{2}(\gamma_0\cos\xi_0-\gamma_1\cos\xi_1).
\end{aligned}
\end{equation}
With no loss of generality, we set the level $|0\rangle$ with a loss rate $\gamma_0=\gamma$ and $|1\rangle$ with a gain rate $\gamma_1=\gamma$, under $\xi_0=-\pi/2$ and $\xi_1=\pi/2$. Then the conditions in Eq.~(\ref{ConditionTwoGene}) can be simplified as
\begin{equation}\label{ConditionNon}
\begin{aligned}
&\Omega(t)=-\frac{2\dot{\theta}(t)+\gamma\sin2\theta(t)}{\sin\left[\varphi+\alpha(t)\right]}\\
&\Delta(t)=\dot{\alpha}(t)+\Omega(t)\cot2\theta(t)\cos\left[\varphi+\alpha(t)\right].
\end{aligned}
\end{equation}
One can verify that the occurrence of EPs places a complex-valued constraint on $\Omega(t)$, which is in sharp contrast to the real-valued function in Eq.~(\ref{ConditionNon}). EPs are found to be avoided by either $\Delta(t)\ne0$ or $\Delta(t)=0$ and $|\Omega(t)|\neq\gamma$.

Using Eq.~(\ref{U0}) with $K=2$, when the system is initially prepared as the state $|\mu_2(t_0)\rangle$, the effective time-evolution operator in the ket space can be written as
\begin{equation}\label{U0two}
U_0(t)=e^{-if_{22}(t)}|\mu_2(t)\rangle\langle\mu_2(t_0)|,
\end{equation}
where the complex phase $f_{22}(t)$ can be calculated in accordance with the definition in Eq.~(\ref{Hamint}) as
\begin{equation}\label{Twof}
\begin{aligned}
f_{22}(t)&=\int_0^tdt'\left[\dot{f}_r(t')+\dot{f}_i(t')\right]\\
&=\int_0^tdt'\left[\dot{f}_d(t')+\dot{f}_g(t')+\dot{f}_i(t')\right].
\end{aligned}
\end{equation}
Here $\dot{f}_r(t)$ and $\dot{f}_i(t)$ indicate the real and imaginary parts of the time derivative of $f_{22}(t)$, respectively. And, $\dot{f}_r(t)$ is composed of the dynamical part $\dot{f}_d(t)$ and the geometric part $\dot{f}_g(t)$. In particular, we have
\begin{equation}\label{RIcomp}
\begin{aligned}
&\dot{f}_d(t)=\frac{1}{2}\Delta(t)\cos2\theta(t)+\frac{1}{2}\Omega(t)\sin2\theta(t)\cos\left[\varphi+\alpha(t)\right],\\
&\dot{f}_g(t)=-\frac{\dot{\alpha}(t)}{2}\cos2\theta(t), \\
&\dot{f}_i(t)=\frac{1}{2}i\gamma\cos2\theta(t).
\end{aligned}
\end{equation}
Under Eq.~(\ref{ConditionNon}), $\dot{f}_r(t)$ can be expressed as $\dot{f}_r(t)=\dot{f}_d(t)+\dot{f}_g(t)=\Omega(t)\cos(\varphi+\alpha)/[2\sin2\theta(t)]$. Equations~(\ref{U0two}) and (\ref{RIcomp}) indicate that the system dynamics is generally not unitary due to the non-Hermitian components in the Hamiltonian~(\ref{Ham}). Nevertheless, the unitarity of the time evolution can be recovered at the final time point under the condition $f_i(t_f)-f_i(t_0)=0$. It means that the final state can be automatically normalized by controlling $\theta(t)$. In addition, the parallel transport condition for non-Hermitian systems~\cite{Dembowski2004Encircling}, i.e., $\dot{f}_d(t)=0$, can be satisfied under the conditions of $\varphi+\alpha(t)=\pi/2$ and $\Delta(t)=0$ due to Eq.~(\ref{RIcomp}). One possible method to acquire a purely geometric phase $f_g(t)$ along the passage $|\mu_2(t)\rangle$ is to set $\alpha(t)$ as a Heaviside step function~\cite{Sjoqvist2012Nonadiabatic} in the time integral~(\ref{Twof}).

In parallel, in the dual space of the system that is controlled by $H^\dagger(t)$, i.e., the Hermitian conjugate of $H(t)$, the conditions in Eq.~(\ref{ConditionNon}) can activate the ancillary basis state $\langle\mu_1(t)|$ as the universal passage. Apparently, when the system is initialized as the state $|\mu_1(t_0)\rangle$, the relevant time-evolution operator $V_0(t)$ is found to be in the same formation as Eq.~(\ref{U0adj}) with $\dot{f}_{11}^*(t)=-\dot{f}_{22}^*(t)$, where $\dot{f}_{22}^*(t)$ is the complex conjugate of $\dot{f}_{22}(t)$ in Eq.~(\ref{Twof}).

To avoid the singularities of $\Delta(t)$ and $\Omega(t)$ in Eq.~(\ref{ConditionNon}), one can alternatively choose $\theta(t)$ and $f_r(t)$ as their variables in practice. Then under the condition in Eq.~(\ref{RIcomp}), Eq.~(\ref{ConditionNon}) can be written as
\begin{subequations}\label{ConditionInv}
\begin{align}
|\Omega(t)|^2&=[2\dot{\theta}(t)+\gamma\sin2\theta(t)]^2+4\dot{f}_r^2(t)\sin^22\theta(t), \label{Omf} \\
\Delta(t)&=\dot{\alpha}(t)+2\dot{f}_r(t)\cos2\theta(t), \label{Deltaf} \\
\dot{\alpha}(t)&=-\frac{\ddot{\theta}\dot{f}_r\sin2\theta-\ddot{f}_r\dot{\theta}\sin2\theta
-2\dot{f}_r\dot{\theta}^2\cos2\theta}{\dot{f}_r^2\sin^22\theta+\dot{\theta}^2}. \label{alphaf}
\end{align}
\end{subequations}
Equation~(\ref{ConditionInv}) allows arbitrary proper functions of $\theta(t)$ and $f_r(t)$. In the presence of small deviations of Rabi frequency and detuning, it was found~\cite{Jin2025ErrCorr} that a sufficiently fast-varying (real-part of) phase can be used to dynamically correct the error. Simply, we can set
\begin{equation}\label{Globalf}
\dot{f}_r(t)=\lambda\dot{\theta}(t)
\end{equation}
with a coefficient $|\lambda|>1$.

\begin{figure}[htbp]
\centering
\includegraphics[width=0.9\linewidth]{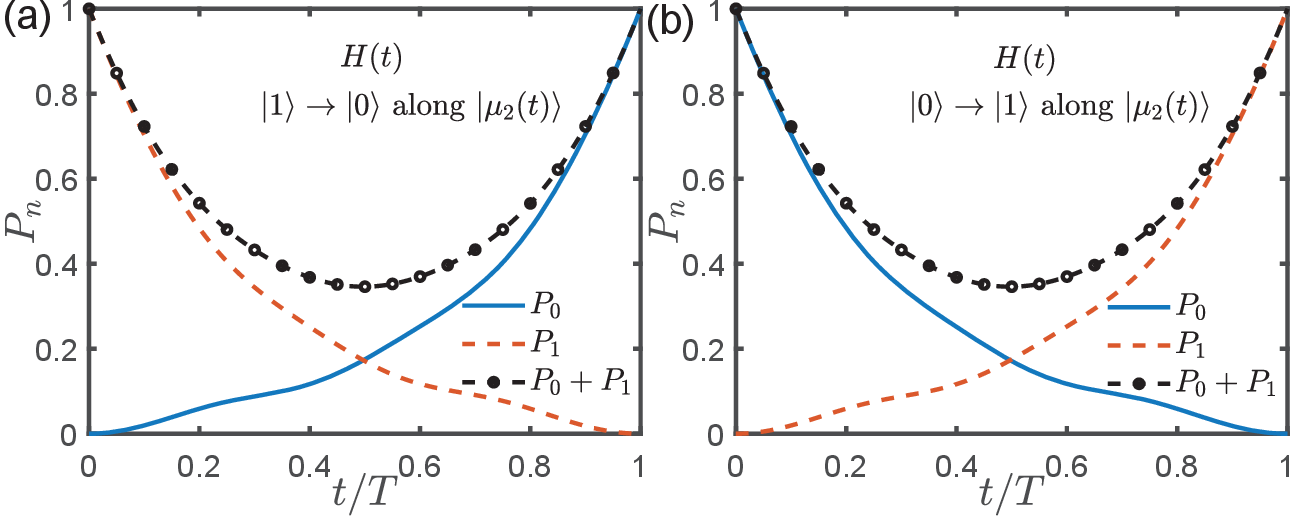}
\includegraphics[width=0.9\linewidth]{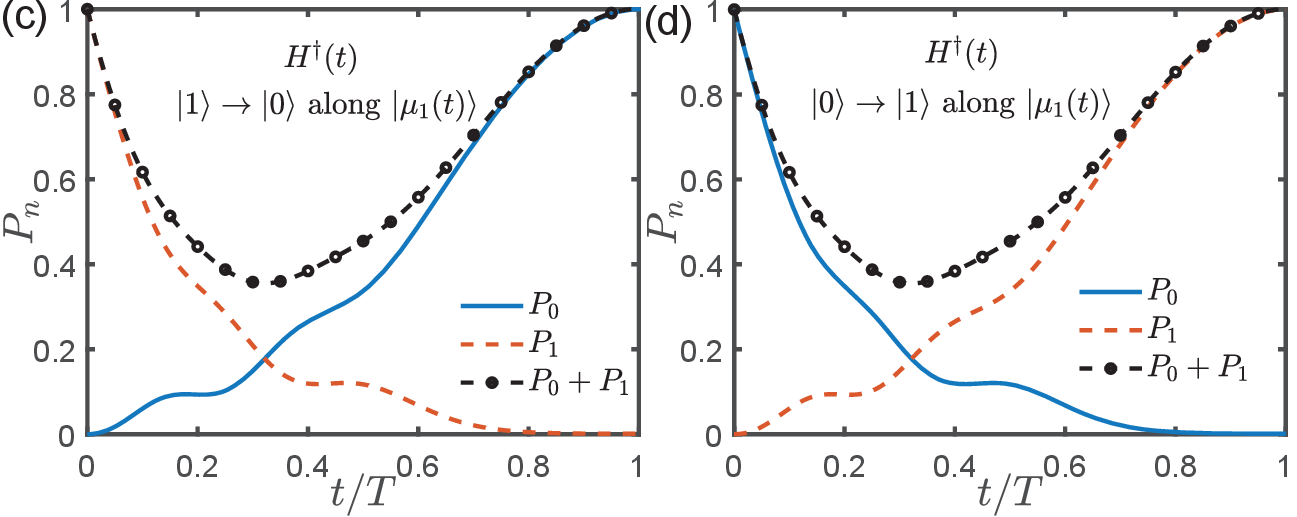}
\caption{Dynamics of the individual population $P_n$ for the atomic levels $|n\rangle$, $n=0,1$, and the full population $P_0+P_1$, in the two-level system under the biothogonal Hamiltonian $H(t)$ in Eq.~(\ref{Ham}) avoiding EPs along the passage $|\mu_{K=2}(t)\rangle$ about the transfer tasks $|1\rangle\rightarrow|0\rangle$ in (a) and $|0\rangle\rightarrow|1\rangle$ in (b), and those in the system controlled by the Hermitian-conjugate Hamiltonian $H^\dagger(t)$ about the transfers (c) $|1\rangle\rightarrow|0\rangle$ and (d) $|0\rangle\rightarrow|1\rangle$ along $|\mu_1(t)\rangle$. $\Omega(t)$, $\Delta(t)$, and $\alpha(t)$ are set by Eqs.~(\ref{Omf}), (\ref{Deltaf}), and (\ref{alphaf}), respectively, under the conditions of $\gamma=|2.1\dot{\theta}(t)|$, $f_r(t)=5\theta(t)$, and (a) $\theta(t)=\pi t/(2T)$, (b) $\theta(t)=-\pi t/(2T)+\pi/2$, (c) $\theta(t)=-(\pi/2)\sin[\pi t/(2T)]$, and (d) $\theta(t)=(\pi/2)\sin[\pi t/(2T)]+\pi/2$. All of them satisfy $f_i(t_f)-f_i(t_0)=0$ with $t_0=0$ and $t_f=T$ and $\Delta(t)\neq0$.}\label{PopuTwo}
\end{figure}

Equation~(\ref{U0two}) [Eq.~(\ref{U0adj})] indicates that there exists at least one useful nonadiabatic passage $|\mu_2(t)\rangle$ [$|\mu_1(t)\rangle$] for the two-level systems controlled by $H(t)$ [$H^\dagger(t)$] in the ket (dual) space. One can alternatively choose $|\mu_2(t)\rangle$ or $|\mu_1(t)\rangle$ [Strictly, it should be $\langle\mu_1(t)|$] for certain quantum tasks, through an appropriate setting about the boundary conditions of $\theta(t)$. Following the passage $|\mu_2(t)\rangle$ in Eq.~(\ref{AnciTwo}), the atomic population can be transferred from $|1\rangle$ to $|0\rangle$ under the boundary conditions of $\theta(t_0)=0$ and $\theta(t_f)=\pi/2$, and from $|0\rangle$ to $|1\rangle$ under the conditions of $\theta(t_0)=\pi/2$ and $\theta(t_f)=0$.

The performance of our protocol that lasts $T$ can be evaluated by the population $P_n\equiv|\langle n|\psi(t)\rangle|^2$, $n=0,1$. Here the pure-state solution $|\psi(t)\rangle$ is obtained by solving the time-dependent Schr\"odinger equation $id/dt|\psi(t)\rangle=H(t)|\psi(t)\rangle$ with the non-Hermitian Hamiltonian $H(t)$ given by Eq.~(\ref{Ham}) or its Hermitian conjugate $H^\dagger(t)$. Figures~\ref{PopuTwo}(a) and \ref{PopuTwo}(b) demonstrate the population dynamics of the system along the passage $|\mu_2(t)\rangle$ in the ket space. Figures~\ref{PopuTwo}(c) and \ref{PopuTwo}(d) demonstrate the population dynamics for the evolution along the passage $\langle\mu_1(t)|$ in the bra space. To show significantly different patterns, the gain or loss rate $\gamma$ is set to be constant in Figs.~\ref{PopuTwo}(a) and \ref{PopuTwo}(b), and time-dependent in Figs.~\ref{PopuTwo}(c) and \ref{PopuTwo}(d). It is straightforward to see that the full population $P_0+P_1$ is not conserved during $t\in(0,T)$, i.e., $P_0(t)+P_1(t)<1$, as indicated by the black dashed line with circles in Fig.~\ref{PopuTwo}. It is a clear manifestation of nonunitary evolution by exchanging population with the environment. Yet the state of system is renormalized at the end of the passage $t=t_f=T$. In other words, the population on the initial state can be transferred to the target state with a unit probability when $t=T$, i.e., $|1\rangle\rightarrow|0\rangle$ and $|0\rangle\rightarrow|1\rangle$ along the passage $|\mu_2(t)\rangle$ in Figs.~\ref{PopuTwo}(a) and \ref{PopuTwo}(b); and $|1\rangle\rightarrow|0\rangle$ and $|0\rangle\rightarrow|1\rangle$ along the passage $|\mu_1(t)\rangle$ in Figs.~\ref{PopuTwo}(c) and \ref{PopuTwo}(d), respectively.

\begin{figure}[htbp]
\centering
\includegraphics[width=0.9\linewidth]{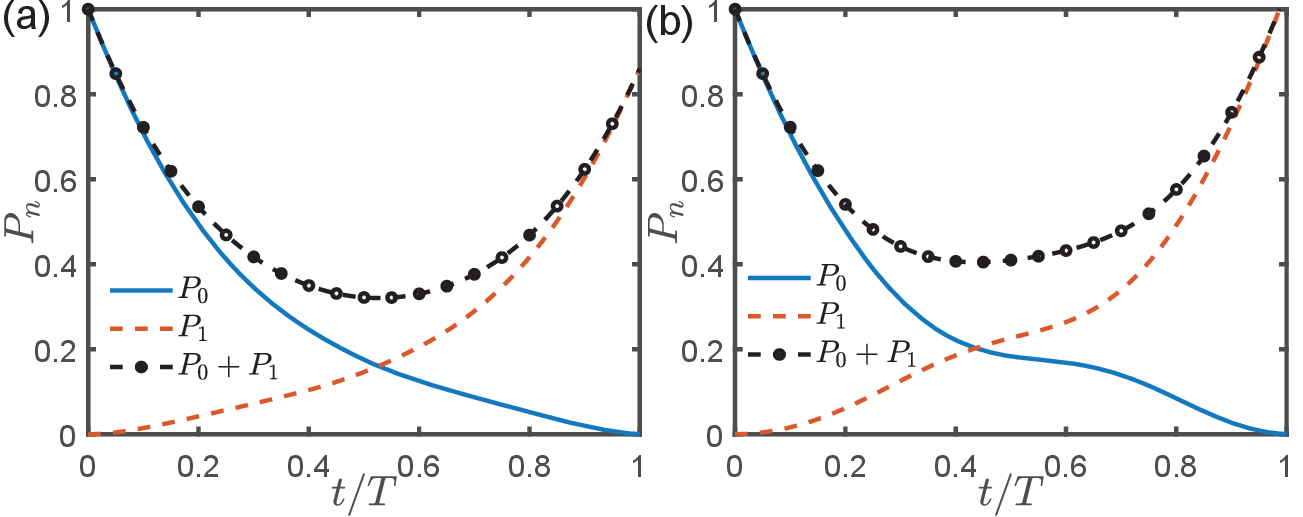}
\caption{Dynamics of the $P_n$ for the atomic levels $|n\rangle$, $n=0,1$, and their sum $P_0+P_1$ in the two-level system controlled by the Hamiltonian~(\ref{Ham}) with parametric deviations in Eq.~(\ref{paraFlu}) along the passage $|\mu_{K=2}(t)\rangle$ about the transfer $|0\rangle\rightarrow|1\rangle$. $\Omega(t)$ and $\Delta(t)$ are deviated as in Eq.~(\ref{paraFlu}) with $\epsilon=-\sin\theta(t)/10$ and $\dot{f}_r(t)$ in Eq.~(\ref{Globalf}) is set with (a) $\lambda=0$ and (b) $\lambda=3$. The other parameters are set the same as Fig.~\ref{PopuTwo}(b).}\label{stable}
\end{figure}

The passage robustness against the parameter fluctuations can be demonstrated in Fig.~\ref{stable} by virtue of our error correction mechanism~(\ref{Globalf}). With no loss of generality, we suppose that $\Omega(t)$ and $\Delta(t)$ in Eq.~(\ref{ConditionInv}) have small deviations as
\begin{equation}\label{paraFlu}
\Omega(t)\rightarrow(1+\epsilon)\Omega(t),\quad \Delta(t)\rightarrow(1+\epsilon)\Delta(t),
\end{equation}
with a perturbative coefficient $\epsilon$. One can verify that its adverse effect can simulate that induced by the initial-state deviations. In Fig.~\ref{stable}(a) without correction, the initial population on the state $|0\rangle$ is transferred to the state $|1\rangle$ with $P_1=0.857$ when $t=T$. In contrast, in Fig.~\ref{stable}(b) with correction, the population on the state $|0\rangle$ can be faithfully transferred to the state $|1\rangle$ at the end of the passage. This result indicates that the instabilities arising from control or initial-state deviations can be mitigated by our triangularization condition in combination with the dynamical correction.

\section{Cyclic population transfer in non-Hermitian three-level systems}\label{NonHermThree}

\subsection{Activation of universal nonadiabatic passages}\label{Activa}

\begin{figure}[htbp]
\centering
\includegraphics[width=0.8\linewidth]{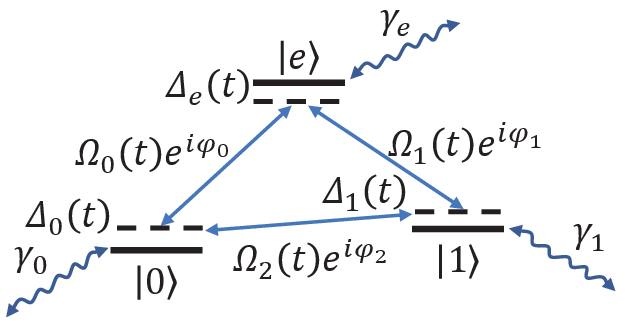}
\caption{Sketch of an open three-level system under control. The transitions $|e\rangle\leftrightarrow|n\rangle$, $n=0,1$, and $|0\rangle\leftrightarrow|1\rangle$ are driven by the respective off-resonant driving fields. The gain or loss rates $\gamma_0$, $\gamma_1$, and $\gamma_e$ are associated with the levels $|0\rangle$, $|1\rangle$, and $|e\rangle$, respectively. }\label{modelThree}
\end{figure}

In this section, our theoretical framework for the non-Hermitian quantum mechanics is examined in an open three-level system, which consists of discrete levels $|0\rangle$, $|1\rangle$, and $|e\rangle$ as shown in Fig.~\ref{modelThree}. The transitions $|e\rangle\leftrightarrow|n\rangle$, $n=0,1$, are driven by the off-resonant laser fields with the detuning $\Delta_e(t)-\Delta_n(t)$, the Rabi frequency $\Omega_n(t)$, and the initial phase $\varphi_n$. The transition $|0\rangle\leftrightarrow|1\rangle$ is driven by the off-resonant laser field with the detuning $\Delta_1(t)-\Delta_0(t)$, the Rabi frequency $\Omega_2(t)$, and the initial phase $\varphi_2$. Additionally, the levels $|0\rangle$, $|1\rangle$, and $|e\rangle$ are assumed with the gain or loss rates $\gamma_0$, $\gamma_1$, and $\gamma_e$, respectively. The whole Hamiltonian can be written as
\begin{equation}\label{HamThree}
\begin{aligned}
H(t)&=\frac{1}{2}\Delta_e(t)|e\rangle\langle e|+\frac{1}{2}\Delta_1(t)|1\rangle\langle1|+\frac{1}{2}\Delta_0(t)|0\rangle\langle0|\\
&+\frac{1}{2}\Big[e^{i\xi_0}\gamma_0|0\rangle\langle0|+e^{i\xi_1}\gamma_1|1\rangle\langle1|\\
&+e^{i\xi_e}\gamma_e|e\rangle\langle e|\Big]+\frac{1}{2}\Big[\Omega_0(t)e^{i\varphi_0}|e\rangle\langle 0|\\
&+\Omega_1(t)e^{i\varphi_1}|e\rangle\langle1|+\Omega_2(t)e^{i\varphi_2}|1\rangle\langle0|+{\rm H.c.}\Big],
\end{aligned}
\end{equation}
where the phases $\xi_0$, $\xi_1$, and $\xi_e$ determine whether the population on the associated levels are under gain or loss. If $|\xi_{n}|\neq\pi/2$, $n=0,1,e$, then they induce the variation in the level splitting in addition to the population gain or loss. Considering the parity operator for the three-level system~\cite{Mandal2021Symmetry}, the Hamiltonian in Eq.~(\ref{HamThree}) exhibits PT symmetry if and only if $\gamma_n\sin\xi_n=0$ for $n=0,1,e$. Any deviation from these conditions results in PT antisymmetry. Again, our protocol is not relevant to that symmetry.

Extending Eq.~(\ref{AnciTwo}) or applying the recipe in Ref.~\cite{Jin2025Entangling}, the ancillary picture used to describe an arbitrary three-level system can be spanned by
\begin{equation}\label{AnciThree}
\begin{aligned}
|\mu_1(t)\rangle&=\cos\theta(t)e^{i\frac{\alpha(t)}{2}}|0\rangle-\sin\theta(t)e^{-i\frac{\alpha(t)}{2}}|1\rangle,\\
|\mu_2(t)\rangle&=\cos\phi(t)e^{i\frac{\beta(t)}{2}}|b(t)\rangle-\sin\phi(t)e^{-i\frac{\beta(t)}{2}}|e\rangle,\\
|\mu_3(t)\rangle&=\sin\phi(t)e^{i\frac{\beta(t)}{2}}|b(t)\rangle+\cos\phi(t)e^{-i\frac{\beta(t)}{2}}|e\rangle,
\end{aligned}
\end{equation}
where the superposed state defined as $|b(t)\rangle\equiv\sin\theta(t)\exp[i\alpha(t)/2]|0\rangle+\cos\theta(t)\exp[-i\alpha(t)/2]|1\rangle$ is orthonormal to $|\mu_1(t)\rangle$, and $|\mu_2(t)\rangle$ and $|\mu_3(t)\rangle$ constitute another SU(2) generator. The parameters $\theta(t)$ and $\phi(t)$ manipulate the population distribution and $\alpha(t)$ and $\beta(t)$ are relevant to the local phases.

In the Hermitian systems~\cite{Jin2025Entangling}, the ancillary basis states can be activated as the useful nonadiabatic passages, when the Hamiltonian contains the terms describing the Rabi oscillation between the states $|0\rangle$ and $|1\rangle$, and those between the states $|e\rangle$ and $|b_1(t)\rangle$. In this spirit, we consider the detunings in the non-Hermitian Hamiltonian (\ref{HamThree}) as
\begin{equation}\label{ConditionDet}
\begin{aligned}
\Delta_1(t)&=-\Delta_e(t)\cos^2\theta(t)+\Delta(t),\\
\Delta_0(t)&=-\Delta_e(t)\sin^2\theta(t)-\Delta(t),\\
\end{aligned}
\end{equation}
where $\Delta(t)$ is the scaling detuning, and the Rabi-frequencies as
\begin{equation}\label{ConditionThree}
\begin{aligned}
\Omega_0(t)e^{i\varphi_0}&=\Omega(t)\sin\theta(t)e^{i\left[\varphi-\frac{\alpha(t)}{2}\right]},\\
\Omega_1(t)e^{i\varphi_1}&=\Omega(t)\cos\theta(t)e^{i\left[\varphi+\frac{\alpha(t)}{2}\right]},\\
\Omega_2(t)e^{i\varphi_2}&=\Omega_a(t)e^{i\varphi_a}-\frac{1}{2}\Delta_e(t)\sin2\theta(t)e^{-i\alpha(t)},
\end{aligned}
\end{equation}
with the scaling Rabi frequencies $\Omega(t)$ and $\Omega_a(t)$. Under the conditions in Eqs.~(\ref{ConditionDet}) and (\ref{ConditionThree}), $H(t)$ in Eq.~(\ref{HamThree}) can be transformed as
\begin{equation}\label{HamThreeSuff}
\begin{aligned}
&H(t)=\frac{1}{2}\Delta_e(t)\left[|e\rangle\langle e|-|b_1(t)\rangle\langle b_1(t)|\right]+\frac{1}{2}\Delta(t)(|1\rangle\langle1|\\
&-|0\rangle\langle0|)+\frac{1}{2}\left[e^{i\xi_0}\gamma_0|0\rangle\langle0|
+e^{i\xi_1}\gamma_1|1\rangle\langle1|+e^{i\xi_e}\gamma_e|e\rangle\langle e|\right]\\
&+\frac{1}{2}\left[\Omega(t)e^{i\varphi}|e\rangle\langle b_1(t)|+\Omega_a(t)e^{i\varphi_a}|1\rangle\langle0|+{\rm H.c.}\right].
\end{aligned}
\end{equation}

Substituting the ancillary basis states in Eq.~(\ref{AnciThree}) to the triangularization condition in Eq.~(\ref{Tri}) with the non-Hermitian Hamiltonian~(\ref{HamThreeSuff}), we have
\begin{equation}\label{ConditionThreeOm}
\begin{aligned}
&\Omega(t)=\Big[-(\gamma_0\sin\xi_0\sin^2\theta+\gamma_1\sin\xi_1\cos^2\theta\\
&+\gamma_e\sin\xi_e)\sin\phi\cos\phi-2\dot{\phi}\Big]/\sin(\varphi+\beta),\\
&2\Delta_e(t)=2\dot{\beta}(t)+\Delta\cos2\theta-\dot{\alpha}\cos2\theta\\
&+\gamma_0\cos\xi_0\sin^2\theta+\gamma_1\cos\xi_1\cos^2\theta-\gamma_e\cos\xi_e\\
&+\Omega_a\sin2\theta\cos(\varphi_a+\alpha)+2\Omega\cot2\phi\cos(\varphi+\beta),\\
\end{aligned}
\end{equation}
where $\Omega_a(t)$, $\Delta(t)$ and $\dot{\alpha}(t)$ take the similar formation as Eq.~(\ref{ConditionTwoGene}) under the replacements of $\Omega(t)\rightarrow\Omega_a(t)$, and $\varphi\rightarrow\varphi_a$.

With no loss of generality, one can assign the loss rate $\gamma_0=\gamma$ to the level $|0\rangle$, the gain rate $\gamma_1=\gamma$ to the level $|1\rangle$, and the gain or loss rate $\gamma_e$ to the level $|e\rangle$, by setting the phases $\xi_0=-\pi/2$ and $\xi_1=\xi_e=\pi/2$. Then the conditions in Eq.~(\ref{ConditionThreeOm}) can be simplified as
\begin{equation}\label{ConditionThreeSim}
\begin{aligned}
\Omega(t)&=-\frac{2\dot{\phi}+(\gamma\cos2\theta+\gamma_e)\sin\phi\cos\phi}{\sin(\varphi+\beta)},\\
\Delta_e(t)&=\dot{\beta}(t)+\frac{\Omega_a}{2\sin2\theta}\cos(\varphi_a+\alpha)\\
&+\Omega\cot2\phi\cos(\varphi+\beta),
\end{aligned}
\end{equation}
and $\Omega_a(t)$ and $\Delta(t)$ adopt the simplified form given in Eq.~(\ref{ConditionNon}) under the preceding replacements.

Using Eq.~(\ref{U0}) with $K=3$, when the system is initially in the state $|\mu_3(t_0)\rangle$, the time-evolution operator in the ket space can be effectively written as
\begin{equation}\label{U0Three}
U_0(t)=e^{-if_{33}(t)}|\mu_3(t)\rangle\langle\mu_3(t_0)|,
\end{equation}
where the complex phase $f_{33}(t)$ can be obtained from the definition in Eq.~(\ref{Hamint}) as
\begin{equation}\label{f33}
f_{33}(t)=\int_0^tdt'\left[\dot{f}_r(t')+\dot{f}_i(t')\right].
\end{equation}
The real and imaginary parts of the time derivative of $\dot{f}_{33}(t)$ can be respectively expressed as
\begin{equation}\label{ReIm}
\begin{aligned}
&\dot{f}_r(t)=\frac{1}{2}\Big[\Delta_e\cos2\phi
+\Delta\sin^2\phi\cos2\theta+\Omega_a\sin^2\phi\sin2\theta\\
&\times\cos(\varphi_a+\alpha)+\Omega\sin2\phi\cos(\varphi+\beta)-(\dot{\alpha}\sin^2\phi\cos2\theta\\
&+\dot{\beta}\cos2\phi)\Big],\\
&\dot{f}_i(t)=\frac{1}{2}i(\gamma\cos2\theta\sin^2\phi+\gamma_e\cos^2\phi).
\end{aligned}
\end{equation}
According to Eq.~(\ref{ConditionNon}), the conditions of $\varphi_a+\alpha(t)=\pi/2$ and $\dot{\alpha}(t)=0$ lead to $\Delta(t)=0$. Then $\dot{f}_r(t)$ in Eq.~(\ref{ReIm}) can be reduced as
\begin{equation}\label{Scalef}
\begin{aligned}
&\dot{f}_r(t)=\frac{1}{2}\Big[\Delta_e\cos2\phi+\Omega_a\sin^2\phi\sin2\theta\cos(\varphi_a+\alpha)\\
&+\Omega\sin2\phi\cos(\varphi+\beta)-\dot{\beta}\cos2\phi\Big]=\frac{\Omega(t)}{2}\frac{\cos(\varphi+\beta)}{\sin2\phi},
\end{aligned}
\end{equation}
where the second equivalence has used the condition in Eq.~(\ref{ConditionThreeSim}). Under the condition of $f_i(t_f)-f_i(t_0)=0$, the system state is automatically normalized at both initial and final time points.

In parallel, the conditions in Eqs.~(\ref{ConditionDet}), (\ref{ConditionThree}), and (\ref{ConditionThreeSim}) can activate the passage $\langle\mu_1(t)|$ in the dual space of the system controlled by $H^\dagger(t)$. When the system is initially prepared as $|\mu_1(t_0)\rangle$, one can find that the relevant time-evolution operator $V_0(t)$ adopts the same formation as Eq.~(\ref{U0adj}). The accumulated phase satisfies $\dot{f}_{11}^*(t)=-\dot{f}_{22}^*(t)$ with $\dot{f}_{22}^*(t)$ given in Eq.~(\ref{Twof}) under the substitution $\Omega(t)\rightarrow\Omega_a(t)$.

To avoid the singularities of $\Omega(t)$ and $\Delta_e(t)$ in Eq.~(\ref{ConditionThreeSim}), we consider the parameters $\theta(t)$, $\phi(t)$, and $f_r(t)$ as independent variables in practical control. Then the conditions in Eq.~(\ref{ConditionThreeSim}) can be transformed as
\begin{equation}\label{ConditionThreeInv}
\begin{aligned}
&|\Omega(t)|^2=[2\dot{\phi}+(\gamma\sin2\theta+\gamma_e)\sin\phi\cos\phi]^2+4\dot{f}_r^2\sin^22\phi,\\
&\Delta_e(t)=\dot{\beta}+2\dot{f}_r(t)\cos2\phi,\\
&\dot{\beta}(t)=-\frac{\ddot{\phi}\dot{f}_r\sin2\phi-\ddot{f}_r\dot{\phi}\sin2\phi
-2\dot{f}_r\dot{\phi}^2\cos2\phi}{\dot{f}_r^2\sin^22\phi+\dot{\phi}^2},
\end{aligned}
\end{equation}
while the parameter $\Omega_a(t)$ takes a similar form to $\Omega(t)$ in Eq.~(\ref{ConditionInv}) under the condition of $\dot{f}_r(t)=0$.

\subsection{Chiral population transfer via universal nonadiabatic passages}\label{Clockwise}

Now we can design a protocol for the chiral/cyclic population transfers in the non-Hermitian three-level system, including the clockwise transfer $|0\rangle\rightarrow|e\rangle\rightarrow|1\rangle\rightarrow|0\rangle$ and the counterclockwise one $|0\rangle\rightarrow|1\rangle\rightarrow|e\rangle\rightarrow|0\rangle$.

A single loop of a clockwise population transfer along the three levels comprises three stages, i.e., $|0\rangle\rightarrow|e\rangle$, $|e\rangle\rightarrow|1\rangle$, and $|1\rangle\rightarrow|0\rangle$. They can be of the equal duration for $T$. Then a completed loop lasts $3T$. In particular, during the first and second stages, the system dynamics is controlled by the Hamiltonian $H(t)$. In other words, the system evolves along the passage $|\mu_3(t)\rangle$ in Eq.~(\ref{AnciThree}) as
\begin{equation}\label{Mu3Sing}
\begin{aligned}
|\mu_3(t)\rangle&=\sin\phi(t)\sin\theta(t)e^{i\frac{\alpha(t)+\beta(t)}{2}}|0\rangle
+\cos\phi(t)e^{-i\frac{\beta(t)}{2}}|e\rangle\\
&+\sin\phi(t)\cos\theta(t)e^{-i\frac{\alpha(t)-\beta(t)}{2}}|1\rangle,
\end{aligned}
\end{equation}
for $t\in[0, 2T]$. When the parameters $\theta(t)$ and $\phi(t)$ satisfy the boundary conditions $\theta(0)=\pi/2$, $\phi(0)=\pi/2$, $\phi(T)=0$, $\theta(2T)=0$, and $\phi(2T)=\pi/2$, the initial population on the state $|0\rangle$ is first transferred to the state $|e\rangle$ at $t=T$, followed by a subsequent transfer to the state $|1\rangle$ along the same passage $|\mu_3(t)\rangle$ when $t=2T$.

To go back from $|1\rangle$ to $|0\rangle$ without involving $|e\rangle$ along a passage available in non-Hermitian quantum mechanics, one can employ the passage $\langle\mu_1(t)|$ in the bra space on the third stage, which is a superposition of merely $|0\rangle$ and $|1\rangle$. That makes the control to be of a concatenated or a hybrid style. Thus during $t\in[2T, 3T]$, the system dynamics is controlled by $H^\dagger(t)$ and it follows the passage $|\mu_1(t)\rangle$ in Eq.~(\ref{AnciThree}) as
\begin{equation}\label{Mu1Sing}
|\mu_1(t)\rangle=\cos\theta(t)e^{i\frac{\alpha(t)}{2}}|0\rangle-\sin\theta(t)e^{-i\frac{\alpha(t)}{2}}|1\rangle.
\end{equation}
Under the boundary conditions $\theta(2T+0^+)=\pi/2$ and $\theta(3T)=\pi$, the population on $|1\rangle$ can be transferred to $|0\rangle$ when $t=3T$. Notably, $|e\rangle$ remains unpopulated throughout stage 3. In contrast, during stage 1 for $|0\rangle\rightarrow|e\rangle$ and stage 2 for $|e\rangle\rightarrow|1\rangle$, $|1\rangle$ and $|0\rangle$ can be temporally populated according to the formation of $|\mu_3(t)\rangle$.

\begin{figure}[htbp]
\centering
\includegraphics[width=0.8\linewidth]{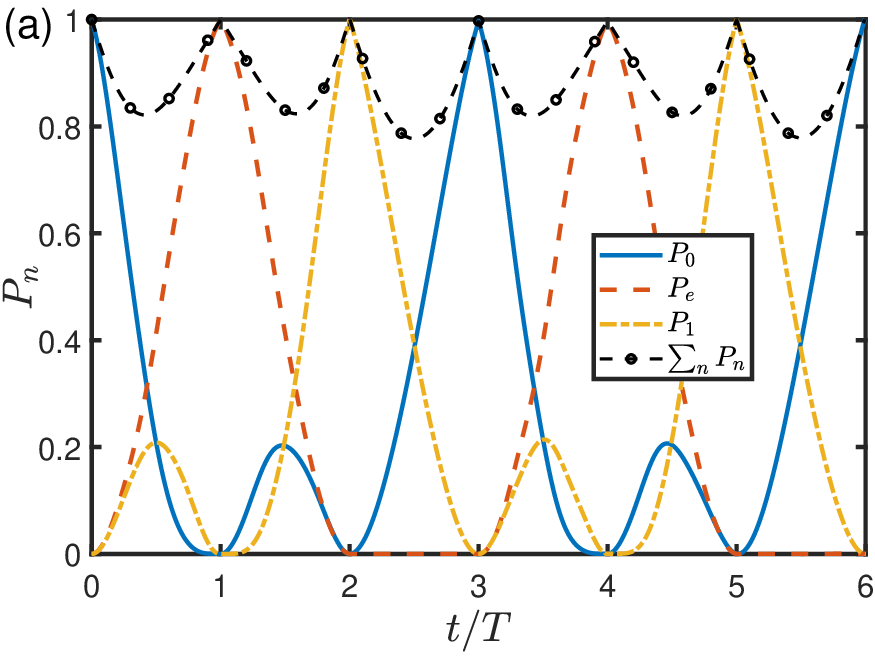}
\includegraphics[width=0.8\linewidth]{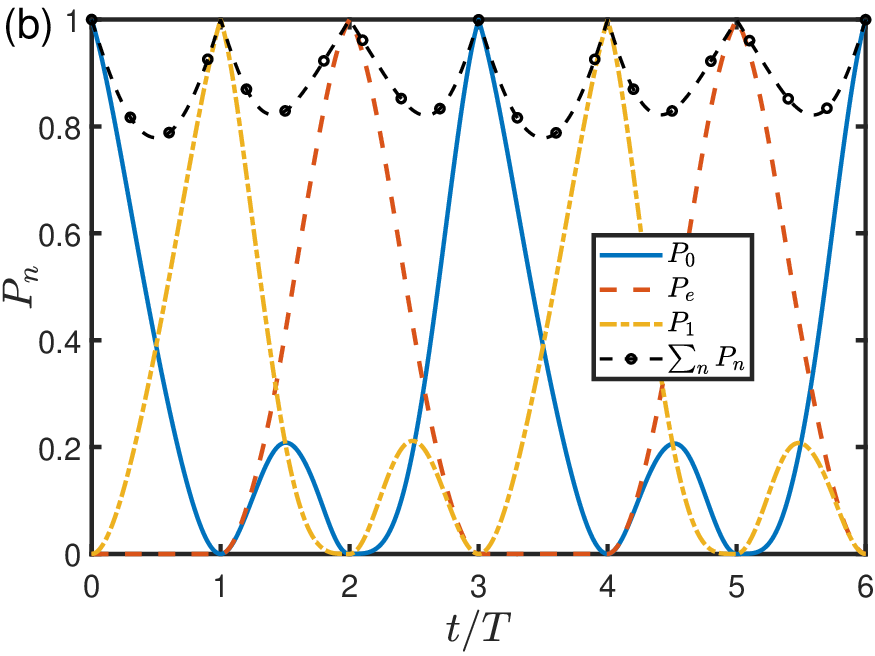}
\caption{Dynamics of the individual population $P_n$ on the levels $|n\rangle$, $n=0,1,e$, and the full population $\sum_nP_n$ for the three-level system, during the first two loops of (a) the clockwise population transfer $|0\rangle\rightarrow|e\rangle\rightarrow|1\rangle\rightarrow|0\rangle$ and (b) the counterclockwise transfer $|0\rangle\rightarrow|1\rangle\rightarrow|e\rangle\rightarrow|0\rangle$. The levels $|0\rangle$, $|1\rangle$, and $|e\rangle$ are assumed with the loss rate $\gamma_0=\gamma$, the gain rate $\gamma_1=\gamma$, and the gain or loss rate $\gamma_e$, respectively, by setting the phases $\xi_0=-\pi/2$ and $\xi_1=\xi_e=\pi/2$. Under the conditions of Eqs.~(\ref{ConditionDet}) and (\ref{ConditionThree}), the parameters $\Delta(t)$, $\Delta_e(t)$, $\Omega(t)$ and $\Omega_a(t)$ are set by Eq.~(\ref{ConditionThreeInv}), where $\gamma=|\dot{\theta}(t)/2|=\pi/(4T)$, $f_r(t)=5\phi(t)$. In (a), $\theta(t)$ and $\phi(t)$ are set by Eq.~(\ref{Cltheta}) and $\gamma_e=\gamma/2$ when $t\in[kT, (k+1)T]$, $k=0,2,3,5$, and $\gamma_e=-\gamma/2$ when $t\in[kT, (k+1)T]$, $k=1,4$; and in (b), $\theta(t)$ and $\phi(t)$ are set by Eq.~(\ref{CounCltheta}) and $\gamma_e$ is set the same as (a).}\label{PopuThree}
\end{figure}

Generally, the $k$th loop of clockwise population transfer during $t\in[3(k-1)T, 3kT]$, $k\geq1$, can be divided into three stages with equal duration $T$. $\theta(t)$ and $\phi(t)$ for each stage can be set as
\begin{subequations}\label{Cltheta}
\begin{align}
\theta(t)&=-\frac{\pi[t-(3k-2)T]}{2T},\quad \phi(t)=\theta(t),\label{period1}\\
\theta(t)&=-\frac{\pi[t-3(k-1)T]}{2T},\quad \phi(t)=\theta(t)-\frac{\pi}{2},\label{period2}\\
\theta(t)&=\frac{\pi[t-(3k-4)T]}{2T},\quad \phi(t)=\theta(t),\label{period3}
\end{align}
\end{subequations}
respectively. The automatical normalization of the non-Hermitian system wavefunction at the beginning and end points of each stage in every loop can be guaranteed as long as the time integral over $\dot{f}_i(t)$ in Eq.~(\ref{ReIm}) vanishes. Under the conditions in Eq.~(\ref{Cltheta}), one can apply the technique of integration by parts to verify that the relevant integral on stages 1, 2, and 3 can be canceled with $\gamma_e=\gamma/2$, $\gamma_e=-\gamma/2$, and $\gamma_e=\gamma/2$, respectively.

Similarly, the $k$th loop of counterclockwise population transfer during $t\in[3(k-1)T, 3kT]$, $k\geq1$, can also be divided into three stages with equal duration. If we employ $\langle\mu_1(t)|$ and $|\mu_3(t)\rangle$ in the first stage and the last two stages, respectively, then the parameters $\theta(t)$ and $\phi(t)$ for each stage can be set as
\begin{subequations}\label{CounCltheta}
\begin{align}
\theta(t)&=-\frac{\pi[t-3(k-1)T]}{2T},\quad \phi(t)=-\theta(t),\label{Counperiod1}\\
\theta(t)&=-\frac{\pi[t-(3k-4)T]}{2T},\quad \phi(t)=-\theta(t)+\pi,\label{Counperiod2}\\
\theta(t)&=\frac{\pi[t-3(k-1)T]}{2T},\quad \phi(t)=-\theta(t)+\pi,\label{Counperiod3}
\end{align}
\end{subequations}
respectively.

Figures~\ref{PopuThree}(a) and \ref{PopuThree}(b) demonstrate the population dynamics for clockwise and counterclockwise population transfers in the open three-level system in Fig.~\ref{modelThree}, respectively. One can find that the full population of the system cannot remain conservative except when $t=kT$, $k$ integer. So, for the most part of time, $P_0+P_1+P_e<1$, as indicated by the black dashed line with circles in Fig.~\ref{PopuThree}. However, the system state is automatically normalized at the end of each evolution stage, i.e., $t=kT$. In Fig.~\ref{PopuThree}(a), the initial population on the state $|0\rangle$ can be transferred to the state $|e\rangle$ with a unit probability when $t=T$, and there exists a slight occupation on the state $|1\rangle$, e.g., $P_1\approx0.21$ when $t=0.6T$. Then it is followed by the transfer to the state $|1\rangle$ with a unit probability when $t=2T$, during which the state $|0\rangle$ is occupied as $P_0\approx0.21$ when $t=1.4T$. During $t\in[2T, 3T]$, there is no occupation on the state $|e\rangle$, and the population on the state $|1\rangle$ can be transferred back to the state $|0\rangle$ with a unit probability when $t=3T$. In Fig.~\ref{PopuThree}(b) for the counterclockwise population transfer, the initial population on the state $|0\rangle$ can be completely transferred to the state $|1\rangle$ when $t=T$, to the state $|e\rangle$ when $t=2T$, and go back to the state $|0\rangle$ when $t=3T$. In both Figs.~\ref{PopuThree}(a) and (b), the following loops can perfectly repeat the same chiral transfer of their respective first loop.

\section{Conclusion}\label{conclusion}

In summary, we propose a theoretical framework to control a general time-dependent non-Hermitian Hamiltonian holding the biorthogonal condition or avoiding exceptional points. Within an ancillary picture, the triangularization condition of the system Hamiltonian is found to be a sufficient condition to obtain a partially solvable time-evolution operator for the non-Hermitian system. It results in at least one individual accessible and universal passage in both ket and bra spaces of the system controlled by a time-dependent biorthogonal Hamiltonian, irrespective of the Hamiltonian symmetry and size. A desired target state can then be faithfully attained through the universal passage and it can be automatically normalized in the end of the control without any artificial normalization in the existing methods. The rapidly varying real part of the global phase can be used to enhance the robustness of the passage against parametric deviation. In theory, the gain and loss rates in our non-Hermitian Hamiltonian can be either time dependent or constant. As illustrative examples, the feasibility of our theory is confirmed by the perfect population transfer in the two-level system and the chiral population transfer in the three-level system.

This work is a nontrivial extension of our universal control framework proposed for Hermitian quantum mechanics. Moreover, the triangularization condition is a generalization of the diagonalization condition supported by the von Neumann equation for the ancillary projectors in case of the Hermitian Hamiltonian. These two conditions thus constitute the foundational elements for our theory of universal quantum control. In principle, our control theory is scalable for many-level and multiple systems as long as a completed set of instantaneous basis states is founded, which applies to both Hermitian and non-Hermitian quantum mechanics.

\section*{Acknowledgment}

We acknowledge grant support from the Science and Technology Program of Zhejiang Province (No. 2025C01028).

\appendix

\section{From D'Alembert principle to universal quantum control}\label{dAlembert}

This Appendix provides a brief review of the universal quantum control theory~\cite{Jin2025Universal} and its methodological connection with the D'Alembert principle in analytical mechanics. We begin with the time-dependent Schr\"odinger equation for a $K$-dimensional quantum system, $id|\psi_m(t)\rangle/dt=H(t)|\psi_m(t)\rangle$, where $|\psi_m(t)\rangle$'s are the pure-state solutions and $H(t)$ is a time-dependent Hermitian Hamiltonian. The system dynamics can be described in the ancillary picture spanned by the instantaneous basis states $|\mu_k(t)\rangle$, $1\le k\le K$. In the rotating frame with respect to $\mathcal{V}(t)\equiv\sum_{k=1}^{K}|\mu_k(t)\rangle\langle\mu_k(t_0)|$, the rotated Hamiltonian $H_{\rm rot}(t)$ takes the same formation as Eq.~(\ref{Hamrot}), except that the dynamical term $\mathcal{H}_{km}(t)$ is Hermitian. In Ref.~\cite{Jin2025Universal}, we proved a theorem that the von Neumann equation for the ancillary projection operator $\Pi_k(t)\equiv|\mu_k(t)\rangle\langle\mu_k(t)|$:
\begin{equation}\label{von}
\frac{d}{dt}\Pi_k(t)=-i\left[H(t), \Pi_k(t)\right],
\end{equation}
is a necessary and sufficient condition for the diagonalization of $H_{\rm rot}(t)$ with a nonvanishing diagonal element in the basis state $|\mu_k(0)\rangle$. In other words, Eq.~(\ref{von}) determines whether $H_{\rm rot}(t)$ can be partially or completely diagonalized. If so for a specific $k$, then the ancillary basis states $|\mu_k(t)\rangle$ can be activated as a universal passage. Consequently, the time-evolution operator in the original picture can be expressed as
\begin{equation}\label{UHerm}
U(t)=\sum_{k=1}^{K'\le K}e^{-if_{kk}(t)}|\mu_k(t)\rangle\langle\mu_k(t_0)|,
\end{equation}
where $K'$ is the number of qualified $|\mu_k(t)\rangle$'s. The global phase $f_{kk}(t)$ takes a similar form as that in Eq.~(\ref{Hamint}). By appropriately choosing the boundary conditions for the passage $|\mu_k(t)\rangle$, it can be applied to the desired quantum task.

To a certain degree, our framework of universal quantum control can be regarded as a quantum version of the D'Alembert principle, which treats the active, constraint, inertial forces on equal footing when dealing with classical dynamics. The equation of motion for the virtual displacement by the D'Alembert principle can be expressed as $(\vec{F}+\vec{R}-\dot{\vec{p}})\cdot\delta\vec{r}=0$, where $\vec{F}$ is the active force, $\vec{R}$ is the constraint force, $\dot{\vec{p}}$ is the inertial force, and $\delta\vec{r}$ is the virtual displacement. When the context of the system dynamics is moved from Newton's law of motion to the Schr\"odinger equation, the active force, the inertial force, the constraint force, and the virtual displacement can be regarded as or enforced by the time-dependent system Hamiltonian $H(t)$, the unitary rotation to the ancillary picture $\mathcal{V}(t)$, the boundary condition of the target state, and the ancillary basis states $|\mu_k(t)\rangle$'s, respectively. As the D'Alembert's principle gives rise to the Lagrange equation, the ancillary projection operator satisfying the von Neumann equation~(\ref{von}) yields the realistic evolution of the system under control in Hermitian quantum mechanics, i.e., Eq.~(\ref{UHerm}).

\section{Reduction from triangularization to diagonalization}\label{appendix}

This Appendix demonstrates that the diagonalization condition for the time-dependent Hermitian Hamiltonian in the ancillary picture spanned by $|\mu_k(t)\rangle$'s, i.e., the von Neumann equation~(\ref{von}) for the projection operator $\Pi_k(t)$, can be obtained from the triangularization condition in Eq.~(\ref{Tri}) when the non-Hermitian Hamiltonian becomes Hermitian~\cite{Jin2025Universal,Jin2025Entangling,Jin2025ErrCorr,Jin2025Rydberg}.

Under the condition of Eq.~(\ref{Tri}) for $1\leq k<K$, we have
\begin{equation}\label{TriSimp}
\begin{aligned}
&\Pi_k(t)\left[\mathcal{H}(t)-\mathcal{A}(t)\right]\sum_{k'>k}^K\Pi_{k'}(t)=0,\\
&\sum_{k'=1}^{k-1}\Pi_{k'}(t)\left[\mathcal{H}(t)-\mathcal{A}(t)\right]\Pi_k(t)=0, \quad k\geq2,
\end{aligned}
\end{equation}
for a specified $k$. If $H(t)=H^\dagger(t)$, then we have
\begin{equation}\label{TriSimpAdd}
\begin{aligned}
&\sum_{k'>k}^K\Pi_{k'}(t)\left[\mathcal{H}(t)-\mathcal{A}(t)\right]\Pi_k(t)=0,\\
&\Pi_k(t)\left[\mathcal{H}(t)-\mathcal{A}(t)\right]\sum_{k'\geq1}^{k-1}\Pi_{k'}(t)=0, \quad k\geq2.
\end{aligned}
\end{equation}
Combining Eqs.~(\ref{TriSimp}) and (\ref{TriSimpAdd}), we obtain
\begin{equation}\label{DifTri}
\left[\mathcal{H}(t)-\mathcal{A}(t)\right]\Pi_k(t)=\Pi_k(t)\left[\mathcal{H}(t)-\mathcal{A}(t)\right],
\end{equation}
which can be expressed in a more compact form:
\begin{equation}\label{Simp}
\left[\mathcal{A}(t), \Pi_k(t)\right]=\left[H(t), \Pi_k(t)\right].
\end{equation}
Using the constraint equation
\begin{equation}\label{Trans}
\mathcal{A}(t)\Pi_k(t)=i|\dot{\mu}_k(t)\rangle\langle\mu_k(t)|,
\end{equation}
Eq.~(\ref{Simp}) can reduce to the von Neumann equation (\ref{von}). It has been proved as a necessary and sufficient condition for the partial diagonalization of $H_{\rm rot}(t)$ in the basis state $|\mu_k(0)\rangle$. When $k$ can run from $1$ to $K$, $H_{\rm rot}(t)$ can be completely diagonalized.

\bibliographystyle{apsrevlong}
\bibliography{ref}

\end{document}